\documentclass[12pt]{article}
\usepackage{graphics,epsfig}
\usepackage{graphicx}

\usepackage{tipa}
\usepackage{stmaryrd}
\usepackage{mathrsfs}
\usepackage{amssymb}
\normalsize
\input{epsf}

\begin{document}
\addtolength{\baselineskip}{.35mm}
\newlength{\extraspace}
\setlength{\extraspace}{2.5mm}
\newlength{\extraspaces}
\setlength{\extraspaces}{2.5mm}

\newcommand{\newsection}[1]{
\vspace{15mm} \pagebreak[3] \addtocounter{section}{1}
\setcounter{subsection}{0} \setcounter{footnote}{0}
\noindent {\Large\bf \thesection. #1} \nopagebreak
\medskip
\nopagebreak}

\newcommand{\newsubsection}[1]{
\vspace{1cm} \pagebreak[3] \addtocounter{subsection}{1}
\addcontentsline{toc}{subsection}{\protect
\numberline{\arabic{section}.\arabic{subsection}}{#1}}
\noindent{\large\bf 
\thesubsection. #1} \nopagebreak \vspace{3mm} \nopagebreak}
\newcommand{\ba}{\begin{eqnarray}
\addtolength{\abovedisplayskip}{\extraspaces}
\addtolength{\belowdisplayskip}{\extraspaces}
\addtolength{\abovedisplayshortskip}{\extraspace}
\addtolength{\belowdisplayshortskip}{\extraspace}}

\newcommand{\be}{\begin{equation}
\addtolength{\abovedisplayskip}{\extraspaces}
\addtolength{\belowdisplayskip}{\extraspaces}
\addtolength{\abovedisplayshortskip}{\extraspace}
\addtolength{\belowdisplayshortskip}{\extraspace}}
\newcommand{\ee}{\end{equation}}
\newcommand{\STr}{{\rm STr}}
\newcommand{\figuur}[3]{
\begin{figure}[t]\begin{center}
\leavevmode\hbox{\epsfxsize=#2 \epsffile{#1.eps}}\\[3mm]
\parbox{15.5cm}{\small
\it #3}
\end{center}
\end{figure}}
\newcommand{\im}{{\rm Im}}
\newcommand{\calm}{{\cal M}}
\newcommand{\call}{{\cal L}}
\newcommand{\sect}[1]{\section{#1}}
\newcommand\hi{{\rm i}}
\def\bea{\begin{eqnarray}}
\def\eea{\end{eqnarray}}

\begin{titlepage}
\begin{center}

\vspace{3.5cm}

{\Large \bf{Shear Viscosity from Effective Couplings of Gravitons}}\\[1.5cm]

{Rong-Gen Cai $^{a,}$\footnote{Email: cairg@itp.ac.cn},} {Zhang-Yu
Nie $^{a,b,}$\footnote{Email: niezy@itp.ac.cn},} {Ya-Wen Sun
$^{a,b,}$\footnote{Email: sunyw@itp.ac.cn},}

\vspace*{0.5cm}

{\it $^{a}$ Institute of Theoretical Physics, \\Chinese Academy of
Sciences P.O.Box 2735, Beijing 100190, China

$^{b}$Graduate University of Chinese Academy of Sciences,
\\ YuQuan
Road 19A, Beijing 100049, China}

\date{\today}
\vspace{3.5cm}

\textbf{Abstract} \vspace{5mm}

\end{center}
We calculate the shear viscosity of field theories with gravity
duals using Kubo-formula by calculating the Green function of dual
transverse gravitons and confirm that the value of the shear
viscosity is fully determined by the effective coupling of
transverse gravitons on the horizon. We calculate the effective
coupling of transverse gravitons for Einstein and Gauss-Bonnet
gravities coupled with matter fields, respectively.  Then we apply
the resulting formula to the case of AdS Gauss-Bonnet gravity with
$F^4$ term corrections of Maxwell field and discuss the effect of
$F^4$ terms on the ratio of the shear viscosity to entropy density.

\end{titlepage}

\newpage
\section{Introduction}
The AdS/CFT
correspondence\cite{Maldacena:1997re,Gubser:1998bc,Witten:1998qj,Aharony:1999ti}
has been a useful tool in the study of properties of strongly
coupled gauge theories. By using AdS/CFT, the shear viscosity of
strongly coupled gauge theories can be calculated in the
hydrodynamic
limit~\cite{Policastro:2001yc,Kovtun:2003wp,Buchel:2003tz,Kovtun:2004de}
on the AdS side. It is found that there is some universality on the
value of the ratio of shear viscosity over entropy density, which is
always $1/4\pi$ in the regimes described by gravity. This ratio is
also conjectured to be a universal lower bound (the KSS
bound~\cite{Kovtun:2003wp}) for all materials. All known materials
in nature by now satisfy this bound. More discussions on the
universality and the bound can be found
in~\cite{{Policastro:2002se},{Policastro:2002tn},{Buchel:2004qq},{Cohen:2007qr},{Son:2007vk},{Cherman:2007fj},
{Chen:2007jq},{Son:2007xw},{Fouxon:2008pz},{Dobado:2008ri},{Landsteiner:2007bd}}.

The universal value of $1/4\pi$ is also obtained in the case with
nonzero chemical potentials turned
on~\cite{Mas:2006dy,{Son:2006em},{Saremi:2006ep},{Maeda:2006by}}.
In~\cite{Cai:2008in}, the value $\eta/s$ is also calculated to be
$1/4\pi$ for
 gauge theories with the gravity dual of
Einstein-Born-Infeld theory. With stringy corrections the value of
$\eta/s$ has a positive derivation from $1/4\pi$ but still satisfies
the KSS
bound~\cite{{Buchel:2004di},Benincasa:2005qc,{Buchel:2008ac},{Buchel:2008wy},
{Buchel:2008sh},{Myers:2008yi},{Buchel:2008ae}}.
However, in\cite{{Brigante:2007nu},{Brigante:2008gz},{KP}} the
authors considered $R^2$ higher derivative gravity corrections and
found that the modification of the ratio of shear viscosity over
entropy density to the conjectured bound is negative, which means
that the lower bound is violated in this condition. The higher
derivative gravity corrections they considered can be seen as
generated from stringy corrections given the vastness of the string
landscape.  A new lower bound, ${4}/{25\pi}$ which is smaller than
$1/4\pi$, is proposed, based on the causality of dual field theory.

In~\cite{Brustein:2008cg,{{Brustein:2008xx}}}, the authors
calculated the ratio of shear viscosity to entropy density for
general gravity theory duals. They identified the value of the ratio
with a quotient of effective couplings~\cite{Brustein:2007jj} of two
different polarizations of gravitons, $\kappa_{xy}$ and
$\kappa_{rt}$ valued on the horizon~\cite{Brustein:2008cg}. This
ratio can be different from $1/4\pi$ in general gravity theories.
In~\cite{Iqbal:2008by}, the authors confirmed the dependence of
shear viscosity on the effective coupling of transverse gravitons
imposed in~\cite{Brustein:2008cg} using the approach of scalar
membrane paradigm. This effective coupling of transverse gravitons
valued on the horizon in general gravity theory may be different
from  the corresponding one in Einstein gravity, which leads to the
value of the ratio different from $1/4\pi$.

In this paper we will first confirm the formula of the dependence of
the shear viscosity on the effective coupling of transverse
gravitons using Kubo formula through a direct calculation of Green
function of transverse gravitons. We reach the same result as using
the membrane paradigm fluid in~\cite{Iqbal:2008by}. The calculation
of the effective coupling of gravitons should be careful because
there are many total derivatives in the effective action of
gravitons, which do not affect the equation of motion of gravitons.
Then we will calculate the effective coupling of transverse
gravitons for Einstein gravity and Gauss-Bonnet gravity coupled to
matter fields separately. In the case of Einstein gravity it would
be easy to show that the value of the ratio is not affected if
matter fields are minimally coupled. However, in the case of
Einstein-Gauss-Bonnet gravity, it has already been observed that the
ratio has corrections if chemical potentials are turned
on~\cite{Ge:2008ni}.  We will also calculate, in the Gauss-Bonnet
gravity, the $F^4$ corrections of Maxwell field to the ratio and
find that the ratio ranges from $(1-4\lambda)/4\pi$ to $1/4\pi$ for
$\varepsilon\geq-{\pi Gl^2}/{72}$. For $\varepsilon<-{\pi
Gl^2}/{72}$, ${(1-4\lambda)}/{4\pi}\leq \eta /s \leq
{(1-4\lambda-\lambda{\pi Gl^2}/{18\varepsilon})}/{4\pi}$, and the
ratio can not reach $1/4\pi$ because the temperature has a lower
bound above zero. Here $\varepsilon$ is a parameter given shortly.

The organization of this paper is as follows. We will first derive
the dependence of shear viscosity on the effective coupling of
transverse gravitons using Kubo formula in Sec.~2. In Sec.~3 we
calculate the effective coupling of transverse gravitons for
Einstein and Gauss-Bonnet gravities coupled with matter fields,
respectively. Then in Sec.~4 we apply the resulting formula of the
dependence of shear viscosity on the effective coupling of
transverse gravitons to AdS Gauss-Bonnet gravity with $F^4$
corrections of Maxwell field. Sec.~5 is devoted to conclusions and
discussions.

\section{The dependence of shear viscosity on the effective coupling}

It was first noted in~\cite{Brustein:2008cg} that the shear
viscosity is determined by the effective coupling of transverse
gravitons. In~\cite{Iqbal:2008by} the authors confirmed this by
using the scalar membrane paradigm fluid. In this section we obtain
this result by calculating the shear viscosity through the energy
momentum/graviton correspondence using the Kubo
formula~\cite{Policastro:2002se,{Son:2007vk}}
\begin{equation}\label{eta}
\eta=\lim_{\omega\rightarrow 0}\frac{1}{2\omega
i}\Big(G^A_{xy,xy}(\omega,0)-G^R_{xy,xy}(\omega,0)\Big),
\end{equation}
where $\eta$ is the shear viscosity, and the retarded Green's
function is defined by
\begin{equation}
G^R_{\mu\nu,\lambda\rho}(k)=-i\int d^4xe^{-ik\cdot x}\theta (t)
\langle[T_{\mu\nu}(x),T_{\lambda\rho}(0)] \rangle.
\end{equation}
These are defined on the field theory side. The advanced Green's
function can be related to the retarded Green's function of energy
momentum tensor by
$G^A_{\mu\nu,\lambda\rho}(k)=G^R_{\mu\nu,\lambda\rho}(k)^{*}$. In
the frame of AdS/CFT correspondence, one is able to compute the
retarded Green's function by making a small perturbation of metric.
Here we choose spatial coordinates so that the momentum of the
perturbation points along the $z$-axis. Then the perturbations can
be written as $h_{\mu\nu}=h_{\mu\nu}(t,z,u)$. In this basis there
are three groups of gravity perturbations, each of which is
decoupled from others: the scalar, vector and tensor
perturbations~\cite{Kovtun:2005ev}. Here we use the simplest one,
the tensor perturbation $h_{xy}$. We use $\phi$ to denote this
perturbation with one index raised $\phi=h^x_y$ and write $\phi$ in
a basis as $\phi(t,u,z)=\phi(u)e^{-i\omega t+ip z}$. To calculate
the Green functions of the energy momentum tensor we should first
fix a background black hole solution and get the equation of motion
for gravitons in this background. In this paper we mainly focus on
the case of Ricci-flat black hole backgrounds. The case for black
holes with hyperbolic horizon topology has been discussed
recently~\cite{{Neupane:2008dc},{Koutsoumbas:2008wy}}. We assume
that the background black hole solution is of the form
\begin{equation}\label{black}
ds^2=-{g(u)(1-u)}dt^2+\frac{1}{h(u)(1-u)}du^2+\frac{r_+^2}{ul^2}(d\vec{x}^2),
\end{equation}
where the horizon of the black hole locates at $u=1$ and the
boundary is at $u=0$, $h(u)$, $g(u)$ are functions of $u$, regular
at $u=1$ and $l$ is the AdS radius which is related to the
cosmological constant by $\Lambda=-6/l^2$. Note here that we impose
the condition $h(u)$ and $g(u)$ are regular at the horizon. This
indicates that the Ricci-flat black hole solution we consider here
should be a non-extremal solution. The calculations below are not
valid for extremal black holes. One can expand the Einstein equation
of motion to the first order of $\phi$ to get the equation of motion
of gravitons, and the effective action of gravitons can be obtained
by expanding the gravity action to the second order of $\phi$. In
the frame of Einstein gravity, the equation of motion of $\phi$ is
just the Klein-Gordon equation for massless scalars. The effective
action for the transverse gravitons is always equal to
\begin{equation}\label{b}
S=\frac{1}{16\pi G}\int d^5x\sqrt{-g}
\bigg(-\frac{1}{2}\bigg)(\nabla_{\mu} \phi\nabla^{\mu}\phi)
\end{equation}
up to some total derivatives in Einstein gravity. However, in
gravity theories with higher derivative corrections, it may not
still be the one for a minimally coupled massless scalar. Now we
consider a specific kind of effective graviton action which is the
same as the one considered in~\cite{Iqbal:2008by}. We write the
effective action in the momentum space
\begin{equation}\label{c}
S=\frac{1}{16\pi G}\int \frac{dwdp}{(2\pi)^2} du\sqrt{-g_0}
\bigg(K(u)\phi'\phi'+w^2K(u)g_{0uu}g_{0}^{00}\phi^2-p^2L(u)\phi^2\bigg)
\end{equation}
up to some total derivatives, where a prime stands for the
derivative with respect to $u$, and
\begin{equation}
\phi(t,u,z)=\int
\frac{dwdp}{(2\pi)^2}\phi(u;k)e^{-iwt+ipz},~~~k=(w,0,0,p),~~~\phi(u;-k)=\phi^*(u;k).
\end{equation}
 This action can be viewed as a
minimally coupled massless scalar with an effective coupling
$K_{eff}(u)=K(u)/g_{0}^{uu}$ plus a $\phi^2$ term proportional to
$p^2$. Here $g_{0\mu\nu}$ denotes the background metric
(\ref{black}). For Einstein gravity, the effective coupling
$K_{eff}=-{1}/{2}$ as can be seen in the action (\ref{b}). However,
in general gravity theories, the effective coupling may depend on
the radial coordinate $u$. In general the effective coupling should
be regular at the horizon, so $1/K(u)$  should have a simple pole at
$u=1$.

 With these assumptions, the equation of motion of the transverse
graviton can be derived from the action (\ref{c})
\begin{equation}\label{a}
\phi''(u,k)+A(u)\phi'(u,k)+B(u)\phi(u,k)=0,
\end{equation}
where
\begin{equation}
A(u)=\frac{(K(u)\sqrt{-g_{0}})'}{K(u)\sqrt{-g_{0}}},
\end{equation}
\begin{equation}
B(u)=-g_{0uu}g_0^{00}w^2+\frac{L(u)}{K(u)}p^2.
\end{equation}
Substituting the metric function yields
\begin{equation}
B(u)=\frac{w^2}{h(u)g(u)(1-u)^2}+\frac{L(u)}{K(u)}p^2.
\end{equation}
Because the shear viscosity only involves physics in the zero
momentum limit, $L(u)$ would not affect the value of $\eta$. The
only constraint on $L(u)$ is that it should be regular at the
horizon $u=1$. In fact we can also have an extra term
$w^2N(u)\phi^2$ in the action (5), and we assume $N(u)$ is also a
function of $u$, which is regular at the horizon $u=1$. The addition
of such a term will not affect the value of $\eta$. Then we follow
the standard procedure to solve this equation (\ref{a}). First we
impose the incoming boundary condition at the horizon so that
\begin{equation}
\phi(u)=(1-u)^{-i\beta w}F(u),
\end{equation}
where $F(u)$ is regular at the horizon. $\beta$ can be calculated by
considering (\ref{a}) in the limit $u\rightarrow1$, which leads to
\begin{equation}\label{beta}
\beta=\frac{1}{\sqrt{h(1)g(1)}}.
\end{equation}
Because we only need to know the $w\rightarrow0$ behavior of this
graviton we can expand the solution in the way
\begin{equation}
F(u)=1+i\beta wF_0(u)+O(w^2)+O(p^2).
\end{equation}
By expanding the equation (\ref{a}) to the first order of $w$, we
get the equation of $F_{0}(u)$
\begin{equation}
F_0''(u)+A(u)F_0'(u)+\frac{1}{(1-u)^2}+\frac{A(u)}{1-u}=0.
\end{equation}
The solution of this linear differential equation can be expressed
as a sum of a specific solution and a general solution. The specific
solution denoted by $F_{0p}(u)$ is easy to find
\begin{equation}
F_{0p}(u)=\ln(1-u).
\end{equation}
The equation for the general solution $F_{0g}$ is
\begin{equation}
F_{0g}''(u)+A(u)F_{0g}'(u)=0.
\end{equation}
Integrating this equation on both sides, we get
\begin{equation}
F_{0g}'(u)=\frac{C}{K(u)\sqrt{-g_{0}}}.
\end{equation}
Further integrating leads to
\begin{equation}\label{e}
F_{0g}(u)=C\int \frac{1}{K(u)\sqrt{-g_0}}du +D,
\end{equation}
where $C$ and $D$ are two integration  constants. From the
assumptions given above we know $F_{0g}'(u)$ should have a simple
pole at $u=1$ because under our assumption of the metric
(\ref{black}), $\sqrt{-g_{0}}$ has no poles and $K(u)$ has a simple
pole at $u=1$. Then if $K(u)\sqrt{-g_{0}}$ is a rational function,
it should have a factor $(1-u)$, so $F_{0g}'(u)$ can be written as a
sum of $b/(1-u)$, where $b$ is a constant, and some function regular
at $u=1$. Thus the integration on $F_{0g}'(u)$ should give
\begin{equation}\label{d}
F_{0g}(u)=b\ln(1-u)+Z(u),
\end{equation}
where $Z(u)$ is a function regular at $u=1$. In many instances,
$K(u)\sqrt{-g_{0}}$ may not be a rational function. For example, in
the case of AdS Born-Infeld black holes~\cite{Cai:2008in}, the
metric function is irrational. In those cases, if we can trust the
Taylor expansions of $K(u)\sqrt{-g_{0}}$ in the region $u\in [0,1]$
to any precision, we still can  have (\ref{d}) as an asymptotic
solution to any desired precision. In this paper, we consider the
cases where (\ref{d}) is valid. We define
$S(u)=K(u)\sqrt{-g_0}/(1-u)$ and $S(1)=\lim_{u\rightarrow 1}
{K(u)\sqrt{-g_0}}/{(1-u)}$, and $S(1)$ should be a finite quantity.
Then by comparing (\ref{d}) and (\ref{e}), we can decide the value
of $s$ and the derivative of the function $Z(u)$. In general, the
solution of $Z(u)$ could not be given explicitly, but fortunately,
only $Z'(u)$ affects the final result. As a result,  we only give
$Z'(u)$ here,
\begin{equation}
b=-\frac{C}{S(1)},
\end{equation}
and
\begin{equation}\label{z}
Z'(u)=\frac{C}{1-u}(\frac{1}{S(u)}-\frac{1}{S(1)}).
\end{equation}
With the specific solution $F_{0p}$ and the general solution
$F_{0g}$, the final solution should be a sum $F_{0p}+jF_{0g}$, where
$j$ is a constant needed to be determined. By requiring  the
solution to be nonsingular at $u=1$, $j$ should be chosen to be
$-1/b$ and $F_{0}(u)$ can be uniquely determined as
\begin{equation}\label{ff}
F_{0}(u)=-\frac{1}{b}Z(u).
\end{equation}
Next we put this solution into (\ref{c}) to give the on-shell
action:
\begin{equation}\label{f}
S_{on-shell}=\frac{1}{16\pi G}\int \frac{dwdp}{(2\pi)^2} du
\bigg((\sqrt{-g}K(u)\phi'\phi)'\bigg).
\end{equation}
Integrating this action gives
\begin{equation}\label{g}
S_{on-shell}=\frac{1}{16\pi G}\int \frac{dwdp}{(2\pi)^2}
\bigg((\sqrt{-g}K(u)\phi'\phi)\bigg)\Big|_{u=1}^{u=0}.
\end{equation}
In the appendix we argue that the other total derivatives in the
bulk action and the Gibbons-Hawking surface term contribution
exactly cancel on the boundary. Thus the resulting effective action
is totally given by the boundary contribution in (\ref{g}).
Following the standard procedure, the retarded Green function can be
calculated as
\begin{equation}
G^R_{xy,xy}(k)=\frac{1}{16\pi
G}2\sqrt{-g_0}K(u){\phi'^*}\phi|_{u=0}.
\end{equation}
Substituting the metric (\ref{black}), the value (\ref{beta}) of
$\beta$, the solution of $\phi$ (\ref{ff}) and (\ref{z}), into
(\ref{eta}), we finally get
\begin{equation}\label{for}
\eta=\frac{1}{16\pi
G}\lim_{w\rightarrow0}\frac{2\sqrt{-g_0}K(u){\phi'^*}\phi|_{u=0}}{iw}=\frac{1}{16\pi
G}\frac{r_+^3}{l^3}\Big(-2K_{eff}(u=1)\Big).
\end{equation}
Thus we arrive at the conclusion that the shear viscosity is fully
determined by the effective coupling of transverse gravitons on the
gravity side. In the Einstein case $K_{eff}=-1/2$ and
$\eta={r_+^3}/{16\pi Gl^3}$, which is the same as the result
obtained in the previous calculations in~\cite{Son:2007vk}. In
gravity theories where the Bekenstein-Hawking entropy area formula
holds, we can further get $\eta/s=(-2K_{eff}(u=1))/4\pi$.

\section{Effective coupling of transverse gravitons}

From the previous section we learn that to calculate the shear
viscosity of a gravity dual, one only needs to know the effective
couplings of the transverse gravitons in this theory.
In~\cite{Brustein:2008cg}, a formula showing effective couplings of
gravitons with different polarizations is given. However, it is not
easy to judge which of the couplings are part of a total derivative
in the general formula. Thus in this section, we calculate the
effective couplings of transverse gravitons separately for Einstein
gravity and Gauss-Bonnet gravity with matter fields minimally
coupled to gravity.

\subsection{For Einstein gravity}

The action of Einstein-Hilbert gravity with matter fields minimally
coupled to the gravity can be written as
\begin{equation}
S=\frac{1}{16\pi G}\int d^5x \sqrt{-g}
\Big(R-2\Lambda+\mathscr{L}_{matter}\Big).
\end{equation}
Here $\mathscr{L}_{matter}$ is the Lagrangian of the matter fields
coupled to gravity which can be the sum of any scalar or gauge
fields.  We assume the background black hole solution is of the form
(\ref{black}), which implies that the matter fields  only depends on
the radial coordinate $u$ and we also assume that
$\mathscr{L}_{matter}$ depends on the metric only through the
coupling of the metric to some ordinary derivatives of matter
fields, such as the cases of minimally coupled scalar field and
Maxwell fields, where covariant derivatives of matter fields are
equivalent to the ordinary derivatives, so that
$\delta_{(2)}\mathscr{L}_{matter}=0$ (see below).  The Einstein
equation of motion for this action is
\begin{equation}\label{eom}
R_{\mu\nu}-\frac{1}{2}g_{\mu\nu}(R-2\Lambda+\mathscr{L}_{matter})+\frac{\delta
\mathscr{L}_{matter}}{\delta g^{\mu\nu}}=0.
\end{equation}
We want to obtain the effective action for the perturbation $h^x_y$.
The effective action for $\phi(u)$ is a sum of two parts: the bulk
action $S_{bulk}$ and the Gibbons-Hawking boundary term $S_{GB}$.
The Gibbons-Hawking term does not affect the effective coupling, and
the bulk effective action should be a sum of a surface contribution
and a term proportional to the equation of motion of $\phi(u)$, the
latter of which vanishes on shell. We derive the effective action by
keeping terms to the second order of $\phi(u)$ in the action:
\begin{equation}
S_{bulk}=\frac{1}{16\pi G}\int d^5 x
\bigg(\delta_{(2)}\sqrt{-g}\Big(R-2\Lambda +\mathscr{L}_{matter}
\Big)+\sqrt{-g}\delta_{(2)}\Big(R-2\Lambda +\mathscr{L}_{matter}
\Big)\bigg).
\end{equation}
Here $\delta_{(2)}(\cdot\cdot\cdot)$ means to only keep terms  of
the second order of $\phi$ in $(\cdot\cdot\cdot)$. We have
$\delta_{(2)}\mathscr{L}_{matter}=0$ because the matter fields only
depend on the radial coordinate $u$ and the metric couples to the
matter fields only through ordinary derivatives. With the $xx$
component of the on-shell Einstein equation of motion
$R-2\Lambda+\mathscr{L}_{matter}=2g^{xx}R_{xx}$, we can get the
action for $\phi(u)$ to be always the form of (\ref{b}) up to some
total derivatives. Thus $K_{eff}=-1/2$ holds in the whole spacetime
and thus of course $K_{eff}=-1/2$ on the horizon. Because the
Bekenstein-Hawking area entropy formula always holds in Einstein
gravity, it is straightforward that the ratio of $\eta/s$ is always
$1/4\pi$ as long as the assumptions in Sec.~2 are satisfied.

\subsection{For Gauss-Bonnet gravity}

In this subsection we calculate the effective coupling of transverse
gravitons for Gauss-Bonnet gravity. We consider the action of
Einstein gravity with Gauss-Bonnet terms as well as matter fields
\begin{equation}
S=\frac{1}{16\pi G}\int d^5 x \sqrt{-g}
\Big(R-2\Lambda+\frac{\lambda l^2}{2}(R^2-4R_{\mu\nu}
R^{\mu\nu}+R_{\mu\nu\rho\sigma}R^{\mu\nu\rho\sigma})+\mathscr{L}_{matter}
\Big).
\end{equation}
The Einstein equation of motion for this action is
\begin{eqnarray} \label{gbeom}
R_{\mu\nu}-\frac{g_{\mu\nu}}{2}\Big(R-2\Lambda+\mathscr{L}_{matter}+\frac{\lambda
l^2}{2}(R^2-4R_{\alpha\beta}
R^{\alpha\beta}+R_{\alpha\beta\rho\sigma}R^{\alpha\beta\rho\sigma})\Big)+\frac{\delta
\mathscr{L}_{matter}}{\delta g^{\mu\nu}} \nonumber \\
+\frac{\lambda
l^2}{2}\Big(2RR_{\mu\nu}-4R_{\rho\mu}{R^\rho}_\nu-4R^{\rho\sigma}R_{\rho\mu\sigma\nu}+2R_{\rho\sigma\lambda\mu}
{R^{\rho\sigma\lambda}}_{\nu}\Big)=0.
\end{eqnarray}
To simplify calculations, we consider a simpler metric
\begin{equation}
\label{in}
ds^2=-\frac{g^{\star}(u)r_+^2}{l^2u}N^2dt^2+\frac{l^2}{4u^2g^{\star}(u)}du^2+\frac{r_+^2}{ul^2}(d\vec{x}^2),
\end{equation}
which is a specific case of (\ref{black}) by setting
$g(u)=\frac{g^{\star}(u)r_+^2}{l^2u(1-u)}N^2$ and
$f(u)=\frac{4u^2g^{\star}(u)}{l^2(1-u)}$, so the calculations in
Sec.~2 are still valid for this metric. $N^2$ is a constant that can
be fixed at the boundary, which is defined in order to make the
solution conformal to flat Minkowski spacetime on the boundary at
$r\to \infty$
\begin{equation}
N^2=\frac{1}{2}(1+\sqrt{1-4\lambda}).
\end{equation}
By keeping the action to the second order of $\phi$ we can get the
effective action for transverse gravitons
\begin{eqnarray}\label{action}
S_{bulk}=\frac{1}{16\pi G}\int d^5x \Big[\delta_{(2)}\sqrt{-g}
\Big(R-2\Lambda+\frac{\lambda l^2}{2}(R^2-4R_{\mu\nu}
R^{\mu\nu}+R_{\mu\nu\rho\sigma}R^{\mu\nu\rho\sigma})+\mathscr{L}_{matter}
\Big) \nonumber \\
+\sqrt{-g}
\delta_{(2)}\Big(R-2\Lambda+\frac{\lambda l^2}{2}(R^2-4R_{\mu\nu}
R^{\mu\nu}+R_{\mu\nu\rho\sigma}R^{\mu\nu\rho\sigma})+\mathscr{L}_{matter}
\Big)\Big].
\end{eqnarray}
Having assumed that the matter fields couple to the metric only
through ordinary derivatives of matter fields, and the matter fields
solution only depends on the radial coordinate $u$, we can see that
the variation $\delta_{(2)}\mathscr{L}_{matter}$ vanishes. Note that
the $xx$ component of the equations of motion (\ref{gbeom})
\begin{eqnarray}\label{onshell}
R_{xx}-\frac{g_{xx}}{2}\Big(R-2\Lambda+\mathscr{L}_{matter}+\frac{\lambda
l^2}{2}(R^2-4R_{\rho\sigma}
R^{\rho\sigma}+R_{\rho\sigma\lambda\theta}R^{\rho\sigma\lambda\theta})\Big)+\frac{\delta
\mathscr{L}_{matter}}{\delta g^{xx}} \nonumber \\ +\frac{\lambda
l^2}{2}\Big(2RR_{xx}-4R_{\rho x}{R^\rho}_x-4R^{\rho\sigma}R_{\rho
x\sigma x}+2R_{\rho\sigma\lambda x}
{R^{\rho\sigma\lambda}}_{x}\Big)=0.
\end{eqnarray}
And ${\delta \mathscr{L}_{matter}}/{\delta g^{xx}}=0$ for the
solution (\ref{in}) we are considering. Substituting the above
equation to (\ref{action}), we find that the effective action for
transverse gravitons can be fully expressed using background metrics
and the derivatives of metrics. Thus we can determine the effective
coupling of the transverse gravitons without knowing the explicit
form of matter fields. The bulk action for transverse graviton
 is therefore
\begin{eqnarray} \label{ac}
S_{bulk}=\frac{1}{16\pi G}\int d^5x \Big[\sqrt{-g}
\delta_{(2)}\Big(R-2\Lambda+\frac{\lambda l^2}{2}(R^2-4R_{\mu\nu}
R^{\mu\nu}+R_{\mu\nu\rho\sigma}R^{\mu\nu\rho\sigma})\Big) \nonumber
\\ +\delta_{(2)}\sqrt{-g} \Big(2{R_{x}}^x+\lambda
l^2(2R{R_{x}}^x-4R_{\rho x}R^{\rho x}-4R^{\rho\sigma}{R_{\rho
x\sigma}}^x+2R_{\rho\sigma\lambda x}R^{\rho\sigma\lambda
x})\Big)\Big].
\end{eqnarray}
Substituting the metric (\ref{in}) into the bulk action, we finally
 get the effective coupling of the transverse graviton $h^x_y$ as
\begin{equation}\label{gauss}
K_{eff}(u)=-\frac{1}{2} \Big(1-2\lambda g^{\star}(u)+2\lambda
ug^{\star \prime}(u) \Big).
\end{equation}
We  see that this effective coupling depends on the background
metric and the first derivative of the metric, and is independent of
explicit form of matter fields. The effect of matter fields is
reflected in the metric function $g^{\star}(u)$. Thus we obtain a
universal formula of the shear viscosity for the AdS Gauss-Bonnet
gravity with arbitrary minimally coupled matter fields, which only
depends on the value of the metric and the first derivative of the
metric on the horizon.

\section{Effects of $F^4$ terms in Gauss-Bonnet theory}

In the case of Einstein gravity, the effective coupling of
transverse gravitons is a constant and not affected by minimally
coupled matter fields. However, for Gauss-Bonnet gravity, the
effective coupling of transverse gravitons (\ref{gauss}) depends on
the value of the metric and its first derivative. Thus when matter
fields are coupled, the value of the ratio $\eta/s$ may be different
from the case of pure Gauss-Bonnet gravity. In \cite{Ge:2008ni},
when Maxwell field is added, the ratio $\eta/s$ gets a positive
correction, compared to the pure AdS Gauss-Bonnet gravity case. Now
we apply the resulting formulas (\ref{for}) and (\ref{gauss}) to the
case of the Gauss-Bonnet-Maxwell theory with $F^4$ terms correction.

The effective action of the theory we are considering is
\cite{Anninos:2008sj}
\begin{eqnarray}
S&=&S_{grav}+S_{matter}\nonumber \\
&=&\frac{1}{16\pi G}\int d^5 x\sqrt{-g}\Big(R-2\Lambda
+\frac{\lambda l^2}{2}
(R^2-4R_{\mu\nu}R_{\mu\nu}+R^{\mu\nu\rho\sigma}R_{\mu\nu\rho\sigma})\Big)\nonumber
\\
 &~&~~~ +\int d^5x\sqrt{-g}\Big(-\frac{1}{4}F_{\mu\nu}F^{\mu\nu}+
c_1(F_{\mu\nu}F^{\mu\nu})^2+c_2F_{\mu\nu}F^{\nu\lambda}F_{\lambda\rho}F^{\rho\mu}\Big),
\end{eqnarray}
where $c_1$ and $c_2$ are two constants and $\Lambda=-6/l^2$. We
consider the Ricci-flat black hole solutions with only $F_{tr}$
component of the Maxwell fields non-vanishing. In this assumption,
the solution  only depends on a combination $\varepsilon$ of $c_1$
and $c_2$~\cite{Anninos:2008sj}, where
\begin{equation}
\varepsilon\equiv 2c_1+c_2.
\end{equation}
The Ricci-flat black hole solution is
\begin{eqnarray} \label{metric}
ds^2&=&-H(r)N^2dt^2+H^{-1}(r)dr^2+\frac{r^2}{l^2}d\vec{x}^2,\nonumber
\\ F_{tr}&=&f(r),
\end{eqnarray}
where
\begin{equation}
H(r)=\frac{r^2}{2\lambda l^2}\bigg(1-\sqrt{1-4\lambda\Big(
1-\frac{ml^2}{r^4}-16\pi G \frac{I(r)l^2}{3r^4}\Big)}\bigg),
\end{equation}
\begin{equation}
I(r)=2\int dr r^3\bigg(\frac{f(r)^2}{4}+3\varepsilon f(r)^4\bigg),
\end{equation}
and $m$ is an integration constant, which is related to the mass of
the black hole solution, $f(r)$ is given by the root of
\begin{equation}\label{eq3}
8\varepsilon f(r)^3+f(r)-\frac{Q}{r^3}=0.
\end{equation}
Here $Q$ is the electric charge of the black hole. The horizon $r_+$
corresponds to the biggest root of $H(r)=0$, that is to say at
$r_+$, one has
\begin{equation}
1-\frac{ml^2}{r_+^4}-16\pi G\frac{I(r_+)l^2}{3r_+^4}=0.
\end{equation}

Before calculating the ratio $\eta/s$, we  first consider the near
boundary behavior of the solution to get the causality constraint
for dual field theory. Although the solution of $f(r)$ and $I(r)$
looks complicated, we can see from (\ref{eq3}) that while
$r\rightarrow \infty$, $f(r)\sim {Q}/{r^3}$, $I \sim -{Q^2}/{4r^3}$.
Then the solution near the boundary becomes the same as the one
without $F^4$ terms~\cite{Ge:2008ni}. As a result, we obtain the
same causality constraint as in \cite{Ge:2008ni}. Following
\cite{Brigante:2008gz,Ge:2008ni}, we can calculate the local ``speed
of graviton"
\begin{equation}
c_g^2(r)
=M_2^2\frac{1-\sqrt{1-4\lambda+M_1}}{2\lambda}(3-2\frac{1-4\lambda+M_2}{1-4\lambda+M_1}),
\end{equation}
where
\begin{equation}
M_1=16\pi G\frac{4\lambda l^2}{3r^4}I+\frac{4\lambda l^2m}{r^4},
\end{equation}
and
\begin{equation}
M_2=16\pi G\frac{2\lambda l^2}{3}(\frac{f^2}{4}+3\varepsilon f^4).
\end{equation}
Near the boundary, $f(r)\sim {Q}/{r^3}$, $I \sim -{Q^2}/{4r^3}$, and
$c^2_g(r)$ can be simplified to be
\begin{equation}
c_g^2(r)-1=(-\frac{5}{2}+\frac{2}{1-4\lambda}-\frac{1}{2\sqrt{1-4\lambda}})\frac{ml^2}{r^4}+\mathcal
{O}(\frac{1}{r^5}).
\end{equation}
With this, we obtain the condition to avoid the causality violation
\begin{equation}
-\frac{5}{2}+\frac{2}{1-4\lambda}-\frac{1}{2\sqrt{1-4\lambda}} <0.
\end{equation}
 This is  the same  result as in
\cite{Brigante:2008gz}, which implies that there is a condition on
the Gauss-Bonnet coefficient $\lambda < 0.09$ in order for the dual
theory to obey the causality law.

Now we turn to the $\eta/s$ ratio. We perform a coordinate
transformation $u=r_+^2/r^2$ on (\ref{metric}), which leads to
\begin{equation}
ds^2=-V(u)N^2dt^2+\frac{r_+^2}{4u^3V(u)}du^2+\frac{r_+^2}{ul^2}d\vec{x}^2,
\end{equation}
where $V(u)$ is just the function obtained by changing variable $r$
in $H(r)$ to $u$. Putting this metric into the formula
(\ref{gauss}),  we have
\begin{equation}
K_{eff}=-\frac{1}{2}(1-4\lambda+2\lambda16\pi
G\frac{I'(1)l^2}{3r_+^4}),
\end{equation}
where
\begin{equation}
I'(1)=-r_+^4\Big(\frac{f(u)^2}{4}+3\varepsilon
f(u)^4\Big)\Big|_{u=1}.
\end{equation}
Note that the area formula of the Bekenstein-Hawking entropy still
holds for Ricci flat black holes in the Gauss-Bonnet
gravity~\cite{Cai}.  We  get the ratio of shear viscosity over
entropy density by inserting the root of (\ref{eq3}) into
(\ref{gauss}) and (\ref{for})
\begin{equation}\label{nos}
\frac{\eta}{s}=-\frac{K_{eff}}{2\pi}=\frac{1}{4\pi}\Big(1-4\lambda[1-\frac{8\pi
Gl^2}{3}(\frac{f_+^2}{4}+3\varepsilon f_+^4)]\Big),
\end{equation}
where $f_+$ denotes $f(u)|_{u=1}$ , which is the root of the cubic
equation (\ref{eq3}) at $r=r_+$.

The temperature of the black hole is easy to calculate as
\begin{equation} \label{Tf}
T= \left.
\frac{1}{2\pi\sqrt{g_{rr}}}\frac{d\sqrt{-g_{tt}}}{dr}\right
|_{r=r_+},
\end{equation}
which gives
\begin{equation} \label{T}
T=\frac{r_+}{\pi l^2}[1-\frac{8\pi
Gl^2}{3}(\frac{f_+^2}{4}+3\varepsilon f_+^4)].
\end{equation}
Then the ratio of $\eta/s$ (\ref{nos}) can be rewritten as
\begin{equation}
\label{in2}
\frac{\eta}{s}=-\frac{K_{eff}}{2\pi}=\frac{1}{4\pi}\Big(1-\frac{4\lambda\pi
l^2}{r_+}T\Big).
\end{equation}
In fact, this relation can also be deduced from the formulas for
$K_{eff}$ (\ref{gauss}) and $T$ (\ref{Tf}). We can see that the
$\eta/s$ ratio depends on the temperature apparently.  As
$T\rightarrow 0$, the corrections to $\eta/s$ vanish. Note that
although the limit $T \to 0$ is well defined in (\ref{in2}), in
fact, some calculations in the above are not valid for extremal
black holes since we start from the metric assumption (\ref{in}) for
a non-extremal black hole.

Now we analyze the correction of $F^4$ term to the ratio of shear
viscosity to entropy density.  To do so, we have to study  the
behavior of the factor $\frac{f_+^2}{4}+3\varepsilon f_+^4$, in
which $f_+$ depends on $r_+$ through the equation (\ref{eq3}). We
define a new function $P(f)$ as
\begin{equation}\label{pf}
P(f)\equiv 8\varepsilon f^3+f=\frac{Q}{r^3}.
\end{equation}
In Figure 1, we plot $P(f)$ as a function of $f$. In the plot, the
red curve denotes $P(f)$ as a function of $f$ in the case of
$\varepsilon>0$, while the blue curve  for the case of
$\varepsilon<0$. We can see from the right hand side of (\ref{pf})
that $P(f)\rightarrow 0$ when $r\rightarrow \infty$. The boundary
condition $f \rightarrow 0$ as $r \rightarrow \infty$ implies that
when $P(f)\rightarrow 0$, $f$ must approach to zero, too. Thus the
physical part of the curve of $P(f)$ should start from the origin
in the figure.
\begin{figure}
\includegraphics[width=13cm] {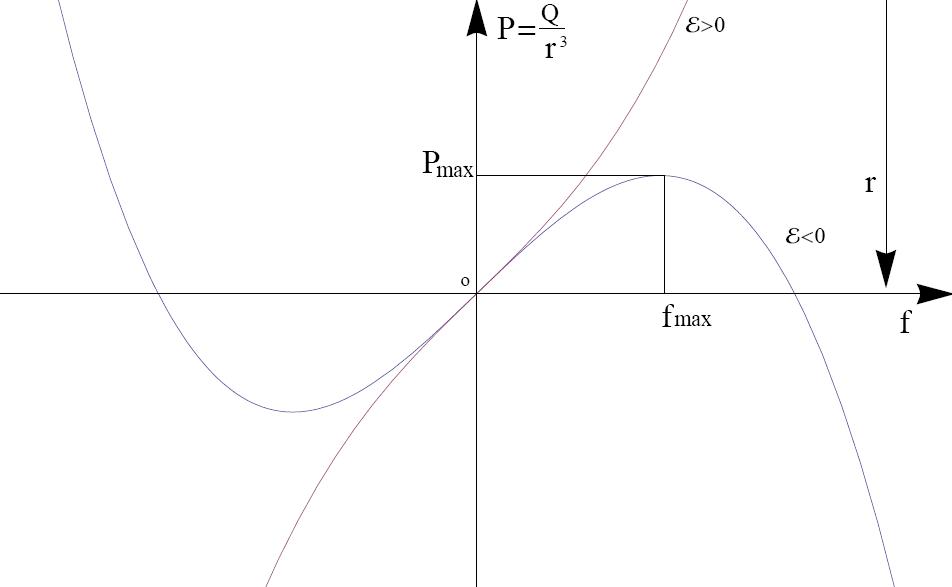}
\caption{\label{pic} $P(f)=8\varepsilon f^3+f$ with
$\varepsilon>0$ and $\varepsilon<0$. The red curve denotes $P(f)$
as a function of $f$ in the case of $\varepsilon>0 $ and the blue
curve for the case of $\varepsilon<0$. The fact that
$P(f)\rightarrow 0$ when $r\rightarrow \infty $ and the boundary
condition $f \rightarrow 0$ as $r \rightarrow \infty$ gives that
when $P(f)\rightarrow 0$, $f$ must approach to zero, too. Thus the
physical part of the curve of $P(f)$ must start from the origin in
the figure. For $Q>0$, the curve of $P(f)$ must be in the above of
$f$-axis. Thus we have $0<f<\infty$ with $\varepsilon>0$ and
$0<f\leq f_{\rm max}$ for $\varepsilon<0$. The behavior of $
P(f_+)$ as a function of $ f_+ $ is the same as $ P(f)$.}
\end{figure}

When $Q>0$, one has $P(f)>0$.  In addition,  $f$ should approach
to $0$ as $r \to \infty$. Therefore, the curves in the region of
$P \ge 0$ and $f\ge 0$ correspond to physical solutions in Figure
1. Namely, in the case of $\varepsilon >0$, the right-hand part of
the red curve is of physical meaning, while in the case of
$\varepsilon<0$,  the blue curve starting from the origin to its
peak is physical.  $P(f)$ is
\begin{equation}\label{rmin}
P_{\rm
max}=\frac{1}{3}\sqrt{\frac{1}{-6\varepsilon}}=\frac{Q}{r_{\rm
min}^3}
\end{equation}
at the peak, where  $f^2=f_{\rm max}^2=-{1}/{24\varepsilon}$. This
implies that in this case, there is a minimal horizon radius
$r_{\rm min}^3= Q\sqrt{-6\varepsilon}/3 $.

For $Q<0$, the situation is similar.  Without loss of generality,
we therefore consider the case of $Q>0$ only.

The case of $\varepsilon>0$ is simple. For extremal black holes,
one has $T=0$, while for large black holes, the temperature
(\ref{Tf}) has the behavior $T=r_+/\pi l^2$ and $f_+ \to 0$.
Therefore, in this case, the ratio is in the range from $1/4\pi$
to $(1-4\lambda)/4\pi$.

 When $\varepsilon<0$, one has $0\leq f_+^2\leq
-{1}/{24\varepsilon}$. In this case, the Hawking temperature is in
the range from $r_+/\pi l^2$ to $r_+(1+\pi Gl^2/72\varepsilon)/\pi
l^2$. In order for the temperature to be positive, one has to impose
the constraint $\varepsilon < - \pi Gl^2/72$. This constraint
excludes the existence of extremal black holes, which requires
$\varepsilon >- \pi G l^2/72$. As a result,  if $\varepsilon < - \pi
Gl^2/72$, the ratio is in the range
\begin{equation}
\frac{1}{4\pi}(1-4\lambda)\leq\frac{\eta}{s}\leq\frac{1}{4\pi}(1-4\lambda-\lambda\frac{\pi
Gl^2}{18\varepsilon}),
\end{equation}
while if $ 0 > \varepsilon >- \pi Gl^2/72$, due to the existence of
extremal black hole, the situation is the same as the case of
$\varepsilon
>0$. Namely, the ratio is in the range
\begin{equation}
\frac{1}{4\pi}(1-4\lambda) \le \frac{\eta}{s} \le  \frac{1}{4\pi}.
\end{equation}

As a result, we see that for arbitrary value of $\varepsilon$, the
effect of correction from $F^2$ and $F^4$ terms is to alleviate the
violation to the universal shear viscosity bound. The ratio ranges
from ${(1-4\lambda)}/{4\pi}$ to ${1}/{4\pi}$ for
$\varepsilon\geq-{\pi Gl^2}/{72}$, and from ${(1-4\lambda)}/{4\pi}$
to ${(1-4\lambda-\lambda{\pi Gl^2}/{18\varepsilon})}/{4\pi}$ for
$\varepsilon<-{\pi Gl^2}/{72}$. The range of the ratio is the same
to the case without $F^4$ terms when $\varepsilon\geq-{\pi
Gl^2}/{72}$. When $\varepsilon<-{\pi Gl^2}/{72}$,  due to the
existence of (non-extremal) minimal black holes whose temperature is
larger than zero, $F^4$ terms lead  the ratio of $\eta/s$ to be
always smaller than ${1}/{4\pi}$.

\section{Conclusions and discussions}

In a general form, we calculated the shear viscosity through AdS/CFT
by calculating the on-shell action of transverse gravitons and
confirmed the argument proposed in \cite{Brustein:2008cg} that the
value of $\eta$ is fully determined by effective couplings of
transverse gravitons on the horizon. Then we calculated the
effective couplings of Einstein gravity and AdS Gauss-Bonnet gravity
with matter fields minimally coupled to the metric separately. We
applied resulting formula for the shear viscosity to the case of AdS
Gauss-Bonnet-Maxwell theory with $F^4$ terms correction of Maxwell
field and found that both $F^4$ terms and the $F^2$ terms together
give a positive $\eta/s$ correction, compared to the case without
Maxwell field. The ratio ranges from $(1-4\lambda)/4\pi$ to
${1}/{4\pi}$ for $\varepsilon\geq-{\pi Gl^2}/{72}$.  When
$\varepsilon<-{\pi Gl^2}/{72}$, the correction makes $\eta/s$ range
from $(1-4\lambda)/4\pi$ to ${(1-4\lambda-\lambda{\pi
Gl^2}/{18\varepsilon})}/{4\pi}$, which is always smaller than
${1}/{4\pi}$.

We have learnt that the universality of $\eta/s=1/4\pi$ is
 valid only for duals of Einstein gravity with arbitrary matter
minimally coupled to gravity. Clearly, in a general gravity theory,
the effective coupling of transverse gravitons on the horizon can be
smaller or bigger than the corresponding value in Einstein gravity.
As a result, the universality of $\eta/s=1/4\pi$ must be violated in
a general gravity theory. So far, most studies have been focused on
the correction to the universal value $1/4\pi$ due to high
derivative terms of gravity, while matter fields are still minimally
coupled to gravity. It would be very interesting to see effect of
non-minimal coupling of matter fields to gravity on the shear
viscosity of dual field theory.

\section*{Appendix}
In this appendix, we  argue that the on shell action on the boundary
is just (\ref{g}) after the Gibbons-Hawking boundary term is
included. We assume that the effective action of gravity is
\begin{equation}
S_{bulk}=\int d^5x\sqrt{-g}\mathcal {L}.
\end{equation}
The variation of this action of gravity with a boundary $\partial M$
is
$$\delta S_{bulk}=\int d^5x\sqrt{-g} G_{\mu\nu}\delta g^{\mu\nu}+\int_{\partial M}d^4x B.$$
Here $B$ is a boundary contribution whose existence originates from
the fact that the derivatives of $\delta g^{\mu\nu}$ are not fixed
to $0$ on the boundary. Then we have to choose a Gibbons-Hawking
term to cancel the contribution of $B$. We assume the
Gibbons-Hawking term to be
\begin{equation}
S_{GB}=\int_{\partial M}d^4x C.
\end{equation}
Then if we choose
\begin{equation}
\delta S_{GB}=\int_{\partial M}d^4x \delta C=-\int_{\partial
M}d^4xB,
\end{equation}
the contribution of $B$ can be eliminated. Here we consider the case
where only $\delta g^{xy}\neq 0$ and we choose $g^{xy}=-g^{xx}\phi$
as in the previous sections. Thus the variation of the action
related to $\delta g^{xy}$ should be
\begin{equation}
\delta S_{bulk}=\int d^5x\sqrt{-g} G_{xy}\delta
g^{xy}+\int_{\partial M}d^4x B(\delta g^{xy}).
\end{equation}
The Gibbons-Hawking term should be designed to eliminate $B$ here,
and this should be valid to any order of $\phi$. To be consistent
with the previous sections, we choose all the functions of $g^{xy}$
here expanded to the second order of $\phi$. Then $G_{xy}=0$ is just
the equation of motion (\ref{a}) for $\phi$. We consider our
effective action (\ref{c}) for $\phi$, and to keep the equation of
motion unaffected,  the full bulk part can be the action (\ref{c})
plus a total derivative
\begin{equation}
S_{bulk}=\frac{1}{16\pi G}\int \frac{dwdp}{(2\pi)^2}
du\Big[\sqrt{-g}
\bigg(K(u)\phi'\phi'+w^2K(u)g_{0uu}g_{0}^{00}\phi^2-p^2L(u)\phi^2\bigg)+G(\phi,\phi')'\Big],
\end{equation}
where $G(\phi,\phi')'$ denotes this total derivative. The variation
of this action is then
\begin{equation}\delta
S_{bulk}=\frac{1}{16\pi G}\int \frac{dwdp}{(2\pi)^2} du
\Big[(EOM)\delta \phi+2(K(u)\sqrt{-g}\phi'\delta\phi)'+(\delta
G(\phi,\phi'))'\Big].
\end{equation}
Because we have $\delta \phi=0$ on the boundary, the term
$2(K(u)\sqrt{-g}\phi'\delta\phi)'$, which involves $\delta \phi$,
vanishes on the boundary after using the Stokes theorem.  Thus the
Gibbons-Hawking term $C$ should obey
\begin{equation}
\delta C+\delta G(\phi,\phi')=0
\end{equation}
on the boundary. To simplify the expression we have chosen $16\pi
G=1$. Thus we can choose \begin{equation}C+
G(\phi,\phi')=0.\end{equation} Thus after integration by parts, the
total on-shell effective action becomes a full surface term
\begin{equation}\label{on}
S_{on-shell}=S_{bulk}+S_{GB}=\frac{1}{16\pi G}\int
\frac{dwdp}{(2\pi)^2}\bigg((\sqrt{-g}K(u)\phi'\phi)\bigg)\Big|_{u=1}^{u=0}
\end{equation}
once the Gibbons-Hawking term is included.

In addition, we would  stress  that because we consider at most
two-order derivatives of $\phi$ in the action, $G(\phi,\phi')$ can
be written as a sum of two kinds of terms:
\begin{equation}
G(\phi,\phi')=G_{1}(u)\phi^2+G_2(u)\phi\phi'.
\end{equation}
$\delta (G_{1}(u)\phi^2)$ vanishes on the boundary, so the
Gibbons-Hawking term $C$ would not involve this part. Thus in the
on-shell surface contributions an additional $G_{1}(u)\phi^2$ term
may also exist in (\ref{on}). However, this term only contributes a
real part to the Green function and thus would not affect the value
of $\eta$, so we can ignore this term in the calculations.

\section*{Acknowledgements}
 RGC thanks CQUeST at Sogang University, Korea and APCTP, Pohang, Korea
 for warm hospitality during his visit and COSPA 2008. ZYN would like to thank Bin Hu for kind help and useful discussions.  This work was supported in part by a
grant from Chinese Academy of Sciences, grants from NSFC with No.
10325525 and No. 90403029.


\begin{thebibliography}{99}

\baselineskip 12pt

\bibitem{Maldacena:1997re}
  J.~M.~Maldacena,
  Adv.\ Theor.\ Math.\ Phys.\  {\bf 2}, 231 (1998)
  [Int.\ J.\ Theor.\ Phys.\  {\bf 38}, 1113 (1999)]
  [arXiv:hep-th/9711200].

\bibitem{Gubser:1998bc}
  S.~S.~Gubser, I.~R.~Klebanov and A.~M.~Polyakov,
  Phys.\ Lett.\  B {\bf 428}, 105 (1998)
  [arXiv:hep-th/9802109].

\bibitem{Witten:1998qj}
  E.~Witten,
  Adv.\ Theor.\ Math.\ Phys.\  {\bf 2}, 253 (1998)
  [arXiv:hep-th/9802150].

\bibitem{Aharony:1999ti}
  O.~Aharony, S.~S.~Gubser, J.~M.~Maldacena, H.~Ooguri and Y.~Oz,
  Phys.\ Rept.\  {\bf 323}, 183 (2000)
  [arXiv:hep-th/9905111].

\bibitem{Policastro:2001yc}
  G.~Policastro, D.~T.~Son and A.~O.~Starinets,
  Phys.\ Rev.\ Lett.\  {\bf 87}, 081601 (2001)
  [arXiv:hep-th/0104066].

\bibitem{Kovtun:2003wp}
  P.~Kovtun, D.~T.~Son and A.~O.~Starinets,
  JHEP {\bf 0310}, 064 (2003)
  [arXiv:hep-th/0309213].

\bibitem{Buchel:2003tz}
  A.~Buchel and J.~T.~Liu,
  Phys.\ Rev.\ Lett.\  {\bf 93}, 090602 (2004)
  [arXiv:hep-th/0311175].

\bibitem{Kovtun:2004de}
  P.~Kovtun, D.~T.~Son and A.~O.~Starinets,
  Phys.\ Rev.\ Lett.\  {\bf 94}, 111601 (2005)
  [arXiv:hep-th/0405231].
\bibitem{Policastro:2002se}
  G.~Policastro, D.~T.~Son and A.~O.~Starinets,
  JHEP {\bf 0209}, 043 (2002)
  [arXiv:hep-th/0205052].

\bibitem{Policastro:2002tn}
  G.~Policastro, D.~T.~Son and A.~O.~Starinets,
  JHEP {\bf 0212}, 054 (2002)
  [arXiv:hep-th/0210220].
\bibitem{Buchel:2004qq}
  A.~Buchel,
  Phys.\ Lett.\  B {\bf 609}, 392 (2005)
  [arXiv:hep-th/0408095].










\bibitem{Cohen:2007qr}
  T.~D.~Cohen,
  Phys.\ Rev.\ Lett.\  {\bf 99}, 021602 (2007)
  [arXiv:hep-th/0702136].
\bibitem{Son:2007vk}
  D.~T.~Son and A.~O.~Starinets,
  Ann.\ Rev.\ Nucl.\ Part.\ Sci.\  {\bf 57}, 95 (2007)
  [arXiv:0704.0240 [hep-th]].

\bibitem{Cherman:2007fj}
  A.~Cherman, T.~D.~Cohen and P.~M.~Hohler,
  JHEP {\bf 0802}, 026 (2008)
  [arXiv:0708.4201 [hep-th]].

\bibitem{Chen:2007jq}
  J.~W.~Chen, M.~Huang, Y.~H.~Li, E.~Nakano and D.~L.~Yang,
  arXiv:0709.3434 [hep-ph].

\bibitem{Son:2007xw}
  D.~T.~Son,
  Phys.\ Rev.\ Lett.\  {\bf 100}, 029101 (2008)
  [arXiv:0709.4651 [hep-th]].

\bibitem{Fouxon:2008pz}
  I.~Fouxon, G.~Betschart and J.~D.~Bekenstein,
  Phys.\ Rev.\  D {\bf 77} (2008) 024016
  [arXiv:0710.1429 [gr-qc]].

\bibitem{Dobado:2008ri}
  A.~Dobado, F.~J.~Llanes-Estrada and J.~M.~T.~Rincon,
  arXiv:0804.2601 [hep-ph].

\bibitem{Landsteiner:2007bd}
  K.~Landsteiner and J.~Mas,
  JHEP {\bf 0707}, 088 (2007)
  [arXiv:0706.0411 [hep-th]].











\bibitem{Mas:2006dy}
  J.~Mas,
  JHEP {\bf 0603}, 016 (2006)
  [arXiv:hep-th/0601144].

\bibitem{Son:2006em}
  D.~T.~Son and A.~O.~Starinets,
  JHEP {\bf 0603}, 052 (2006)
  [arXiv:hep-th/0601157].

\bibitem{Saremi:2006ep}
  O.~Saremi,
  JHEP {\bf 0610}, 083 (2006)
  [arXiv:hep-th/0601159].

\bibitem{Maeda:2006by}
  K.~Maeda, M.~Natsuume and T.~Okamura,
  Phys.\ Rev.\  D {\bf 73}, 066013 (2006)
  [arXiv:hep-th/0602010].


\bibitem{Cai:2008in}
  R.~G.~Cai and Y.~W.~Sun,
  JHEP {\bf 0809}, 115 (2008)
  [arXiv:0807.2377 [hep-th]].
\bibitem{Buchel:2004di}
  A.~Buchel, J.~T.~Liu and A.~O.~Starinets,
  Nucl.\ Phys.\  B {\bf 707}, 56 (2005)
  [arXiv:hep-th/0406264].

\bibitem{Benincasa:2005qc}
  P.~Benincasa and A.~Buchel,
  JHEP {\bf 0601}, 103 (2006)
  [arXiv:hep-th/0510041].

\bibitem{Buchel:2008ac}
  A.~Buchel,
  arXiv:0801.4421 [hep-th].

\bibitem{Buchel:2008wy}
  A.~Buchel,
  arXiv:0804.3161 [hep-th].

\bibitem{Buchel:2008sh}
  A.~Buchel,
  arXiv:0805.2683 [hep-th].

\bibitem{Myers:2008yi}
  R.~C.~Myers, M.~F.~Paulos and A.~Sinha,
  arXiv:0806.2156 [hep-th].


\bibitem{Buchel:2008ae}
  A.~Buchel, R.~C.~Myers, M.~F.~Paulos and A.~Sinha,
  arXiv:0808.1837 [hep-th].


\bibitem{Brigante:2007nu}
  M.~Brigante, H.~Liu, R.~C.~Myers, S.~Shenker and S.~Yaida,
  Phys.\ Rev.\  D {\bf 77}, 126006 (2008)
  [arXiv:0712.0805 [hep-th]];

\bibitem{Brigante:2008gz}
  M.~Brigante, H.~Liu, R.~C.~Myers, S.~Shenker and S.~Yaida,
  Phys.\ Rev.\ Lett.\  {\bf 100}, 191601 (2008)
  [arXiv:0802.3318 [hep-th]].


\bibitem{KP}  Y.~Kats and P.~Petrov,
  arXiv:0712.0743 [hep-th].
\bibitem{Brustein:2008cg}
  R.~Brustein and A.~J.~M.~Medved,
  arXiv:0808.3498 [hep-th].
\bibitem{Brustein:2008xx}
  R.~Brustein and A.~J.~M.~Medved,
  arXiv:0810.2193 [hep-th].
\bibitem{Brustein:2007jj}
  R.~Brustein, D.~Gorbonos and M.~Hadad,
  arXiv:0712.3206 [hep-th].


\bibitem{Iqbal:2008by}
  N.~Iqbal and H.~Liu,
  arXiv:0809.3808 [hep-th].

\bibitem{Ge:2008ni}
  X.~H.~Ge, Y.~Matsuo, F.~W.~Shu, S.~J.~Sin and T.~Tsukioka,
  JHEP {\bf 0810}, 009 (2008)
  [arXiv:0808.2354 [hep-th]].

\bibitem{Kovtun:2005ev}
  P.~K.~Kovtun and A.~O.~Starinets,
  Phys.\ Rev.\  D {\bf 72}, 086009 (2005)
  [arXiv:hep-th/0506184].

\bibitem{Neupane:2008dc}
I.~P.~Neupane and N.~Dadhich,
arXiv:0808.1919 [hep-th].

\bibitem{Koutsoumbas:2008wy}
G.~Koutsoumbas, E.~Papantonopoulos and G.~Siopsis,
Horizons,'' arXiv:0809.3388 [hep-th].

\bibitem{Anninos:2008sj}
  D.~Anninos and G.~Pastras,
  arXiv:0807.3478 [hep-th].

\bibitem{Cai}
R.~G.~Cai,
  Phys.\ Rev.\  D {\bf 65}, 084014 (2002)
  [arXiv:hep-th/0109133];
    R.~G.~Cai and Q.~Guo,
  Phys.\ Rev.\  D {\bf 69}, 104025 (2004)
  [arXiv:hep-th/0311020];
 R.~G.~Cai,
  Phys.\ Lett.\  B {\bf 582}, 237 (2004)
  [arXiv:hep-th/0311240].





\end{thebibliography}
\end{document}